\documentstyle[preprint,aps]{revtex}
\begin{document}
\draft
\title
{Bosonization of the Low Energy Excitations of Fermi Liquids}
\author{A.~H.~Castro Neto and Eduardo Fradkin}
\bigskip
%
\address
{Loomis Laboratory of Physics\\
University of Illinois at Urbana-Champaign\\
1100 W.Green St., Urbana, IL, 61801-3080}
%
\maketitle

\begin{abstract}
We bosonize the low energy excitations of Fermi Liquids in
any number of dimensions in the limit of long wavelengths.
The bosons are coherent superposition of electron-hole pairs
and are related with the displacements of the Fermi Surface
in some arbitrary direction. A coherent-state path integral for the
bosonized theory is derived and it is shown to represent histories of
the shape of the Fermi Surface. The Landau equation for the
sound waves is shown to be exact in the semiclassical approximation for
the bosons. \end{abstract}

\bigskip


\narrowtext

The attempts of description of fermionic systems by bosons date to the
early days of second quantization.
In the early 50's Tomonaga ~\cite{tomonaga}, generalizing earlier work
by Bloch~\cite{bloch} on sound waves in dense fermi systems,  gave an
explicit
construction of the Bloch waves for systems in one
spacial dimension. His work was subsequently generalized by many authors
{}~\cite{lieb} who derived an explicit
fermi-bose transmutation in one-dimensional systems.
These works uncovered deep connections in
relativistic field theories (both fermionic and bosonic) and
with condensed matter systems.

The success of the bosonization approach in one dimension is
related to phase space considerations. Even for non-interacting fermions
two excitations with arbitrarily low energies, moving in the same
direction move at the same speed (the Fermi velocity) and,
hence, are almost a bound state. Consequently, even
the weakest interactions can induce dramatic changes in the nature of
the low lying states. These effects are detected even in perturbation
theory and result in the presence of marginal operators. In dimensions
higher than one, phase space considerations change the physics of the
low-lying states and there are no marginal operators left. This
observation is at the root of the stability of the Landau theory of the
Fermi liquid~\cite{shankar}. It is, thus, hardly surprising
that very few attempts have been made to generalize the bosonization
approach to dimensions higher than one. The earliest serious attempt at
bosonization in higher dimensions was carried out by Luther
{}~\cite{luther2}. Luther constructed a generalized bosonization formula
in terms of the fluctuations of the Fermi Sea along radial directions in
momentum space.
Interest in the construction of bosonized versions of Fermi liquids has
been revived recently in the context of strongly correlated
systems~\cite{god}.
That particle-hole excitations have bosonic character is
well known since the early days of the Landau Theory, {\it e.~g.~}
the sound waves (zero sound collective modes) of the Fermi Surface of
neutral liquids
or plasmons in charged Fermi Liquids~\cite{baym}.
Haldane~\cite{duncan} has recently derived an algebra for the
densities for a Fermi Liquid in a form of a generalized Kac-Moody
algebra (which is central to the construction in one dimension) (see also
ref.~\cite{marston}).

The main point of this article is to derive a description of
the Fermi Liquids as the physics of the dynamics of the Fermi Surface.
The main reason to believe that the Fermi Surface is a {\it real}
dynamical entity
is based on  the following observation:
the total energy of an interacting electronic system  can be
written as an integral in momentum space of the form~\cite{kadanoffbaym}
$E = \sum_{\vec{k}} E_{\vec{k}}$.
For the case of Fermi Liquids, $E_{\vec{k}}$
has a discontinuity at the Fermi Surface (exactly as for the case of
the occupation number ~\cite{luttinger2}). We can think of
this discontinuity as due to
the difference of energy density across the Fermi Surface
{}~\cite{prep}. Hence, we can define an energy per unit area of the Fermi
Surface, {\it i.~e.~} a surface
tension. It readily follows that this surface tension is proportional to
the quasiparticle residue and therefore it vanishes for Non-Fermi
Liquid behavior.
{}From this point of view we see the Fermi Surface as a drumhead
where the elementary excitations are the sound waves (and in
particular the zero sound) which propagate on it. In this paper we will
derive an effective bosonized theory for the dynamics of these
excitations.

Our starting point to approach this problem resembles the
microscopic approaches for the foundations of Fermi Liquid
Theory ~\cite{baym}. However, instead of working with the
dynamics of the response functions, we will work directly with the
dynamics of operators as in the standard procedures of
bosonization in one dimension. Our main result is the derivation of a
bosonized theory of the fluctuations of the shape of the Fermi Surface.
The main ingredients of our construction are  an effective algebra for
the local (in momentum space ) of particle-hole
operators valid in a Hilbert space restricted to the vicinity of the
Fermi Surface and a boson coherent state path-integral constructed from
these states. In particular, we get a bosonized version of Landau's
theory
of the Fermi Liquid in all dimensions, at zero temperature and at very
low energies. The coherent-state path integral can be viewed as a sum
over the histories of the shape of the Fermi Surface, a bosonic {\it
shape field}. For the
case of the Landau hamiltonian we find that resulting bosonic action is
quadratic in the shape fields. We find that the semiclassical
approximation
to our path-integral yields the Landau equation for the sound modes.

For simplicity we will consider a system of interacting
spinless fermions (the spin index can be introduced without
problems in the formalism). The density of fermions at some point
$\vec{r}$ in the
$d$ dimensional space at some time $t$, is given by
\begin{eqnarray}
\rho(\vec{r},t) = \psi^{\dag}(\vec{r},t) \psi(\vec{r},t)
= \sum_{\vec{k},\vec{q}} c^{\dag}_{\vec{k}-\frac{\vec{q}}{2}}(t)
c_{\vec{k}+\frac{\vec{q}}{2}}(t) \, e^{i \vec{q}.\vec{r}}
\end{eqnarray}
where $c^{\dag}_{\vec{k}}$ and $c_{\vec{k}}$ are the creation and
annihilation operator of an electron at some momentum $\vec{k}$
which obey the Fermionic algebra,
$\left\{c^{\dag}_{\vec{k}},c_{\vec{k'}}\right\} =
\delta_{\vec{k},\vec{k'}}$,
where $\{...\}$ is the anticommutator and all other anticommutation
relations are zero.
In the Fermi Liquid theory
{}~\cite{baym} the operator which
appears in the r.h.s. of (1) determines the behavior of the
system.
We will concentrate on this operator which we will denote by
\begin{equation}
n_{\vec{q}}(\vec{k},t)= c^{\dag}_{\vec{k}-\frac{\vec{q}}{2}}(t)
c_{\vec{k}+\frac{\vec{q}}{2}}(t).
\end{equation}
In particular, $n_{0}(\vec{k})$ is the number operator in momentum
space.
As in the standard approaches for the Fermi Liquid theory we
will concentrate our interest in regions close to the Fermi Surface
($\vec{k} \sim \vec{k}_F$) and consider the long wavelength fluctuations
around these regions ($\vec{q} \to 0$).

The equal time commutation relation between the operators defined in (2)
is easily obtained. In the long wavelength limit ($\vec{q} \to 0$) we get
\begin{equation}
\left[n_{\vec{q}}(\vec{k}),n_{\vec{-q'}}(\vec{k'})\right]=
- \delta_{\vec{k},\vec{k'}} \delta_{\vec{q},\vec{q'}} \vec{q}.\nabla(n_{0}
(\vec{k})) + 2 n_{0}(\vec{k}) \delta_{\vec{q},\vec{-q'}} \vec{q}.
\nabla \delta_{\vec{k},\vec{k'}}.
\end{equation}
We are interested in the behavior of systems with a Hilbert space
restricted to the vicinity of the Fermi Surface. We define the Hilbert
space to be the filled Fermi Sea  in the thermodynamic limit $|FS >$ and
the tower
of states obtained by acting finitely on it with local fermion
operators. We will replace
this commutation relation by
a weaker identity valid in the restricted Hilbert space. Furthermore, we
will
make the explicit assumption that we are both in the thermodynamic limit
(the momenta form a continuum) and that the Fermi Surface is {\it
macroscopically large} ($p_F \to \infty$).
 Since the vectors $\vec{k}$
are at the Fermi surface, $\vec{q}$ are very small compared to the the
Fermi momentum, in the restricted Hilbert space it is possible to
replace the  r.~h.~s.~
of (3) by their expectation value in the Filled Fermi Sea $|FS >$ , namely,
\begin{eqnarray}
n_{0}(\vec{k}) \rightarrow \langle n_{0}(\vec{k}) \rangle =
\Theta(\mu-\epsilon_{\vec{k}})
\nonumber \\
\nabla n_{0}(\vec{k}) \rightarrow \nabla \langle n_{0}(\vec{k}) \rangle =
- { \vec{v}}_{\vec{k}} \delta(\mu-\epsilon_{\vec{k}})
\end{eqnarray}
where $\mu$ is the chemical potential, $\epsilon_{\vec{k}}$ is the
one-particle fermion spectrum (from which the Hilbert space is
constructed) and $\vec{v}_{\vec{k}} = \nabla \epsilon_{\vec{k}}$
the velocity of the excitations.
The terms ignored in (3) vanish as $\frac{1}{k_F^2}$ as the size of the
Fermi Surface diverges. These are exactly the same assumptions that
enter in the construction in one-dimension. Notice that the state $|FS
>$, which is used to normal-order the operators, is not necessarily the
ground state of the system of interest. However, this approach will
succeed only the true ground state belongs to the restricted space.

Hence, within this approximation the commutators of the operators
$n_{\vec{q}}(\vec{k})$ become c-numbers, namely,
\begin{equation}
\left[n_{\vec{q}}(\vec{k}),n_{\vec{-q'}}(\vec{k'})\right]=
\delta_{\vec{k},\vec{k'}} \delta_{\vec{q},\vec{q'}} \vec{q}.\vec{v}_{\vec{k}}
\delta(\mu-\epsilon_{\vec{k}}).
\end{equation}
where we drop the last term in the r.h.s. of (3) because its matrix
elements near the Fermi Surface are down by powers of $\frac{1}{k_F}$
(the leading term being zero).
We now define the operator
\begin{equation}
a_{\vec{q}}(\vec{k}) = n_{\vec{q}}(\vec{k})
\Theta(sgn(q)) + n_{-\vec{q}}(\vec{k})\Theta(-sgn(q))
\end{equation}
where the $sgn(q)$ is $+1$ if $\vec{q}$ is outside the Fermi Surface and
$-1$ if it is inside the Fermi Surface. Observe that (6) is not well
defined at $q=0$ which requires special care. We will come back to
this issue below.
The adjoint of (6) is simply,
\begin{equation}
a^{\dag}_{\vec{q}}(\vec{k}) =
n_{-\vec{q}}(\vec{k}) \Theta(sgn(q)) + n_{\vec{q}}(\vec{k})\Theta(-sgn(q))
\end{equation}
where we used, from the definition (2), that
$n^{\dag}_{\vec{q}}(\vec{k}) = n_{-\vec{q}}(\vec{k})$.

It is important to notice that the operator defined in (6) annihilates
the filled Fermi Sea
\begin{equation}
a_{\vec{q}}(\vec{k}) \mid FS \rangle = 0.
\end{equation}
Moreover, from the commutation relations (5) we easily obtain,
\begin{equation}
\left[a_{\vec{q}}(\vec{k}),a^{\dag}_{\vec{q'}}(\vec{k'})\right]=
\mid \vec{q}.\vec{v}_{\vec{k}} \mid  \delta(\mu-\epsilon_{\vec{k}})
\delta_{\vec{k},\vec{k'}} \left(\delta_{\vec{q},\vec{q'}}
+\delta_{\vec{q},-\vec{q'}}\right),
\end{equation}
all other commutators vanish.

Eq. (8) and (9) show that the elementary excitations have bosonic
character,
electron-hole pairs, and are created and annihilated close to the
Fermi Surface by these operators, moreover, they span the Hilbert
space of low energy. But we need an interpretation for these excitations.
It is natural now to define the coherent state~\cite{stone}
\begin{equation}
\mid u_{\vec{q}}(\vec{k}) \rangle = U (\vec{k}) \mid FS \rangle
\end{equation}
where
\begin{equation}
U(\vec{k}) = \exp\left(\sum_{\vec{q}} \frac{v_{\vec{k}}}
{2 \mid \vec{q}.\vec{v}_{\vec{k}} \mid} \, u_{\vec{q}}(\vec{k}) \,
a^{\dag}_{\vec{q}}(\vec{k})\right)
\end{equation}
where the sum (and all other sums that follow are restricted to be close to
$q \to 0$). Observe that from the definition (7) we have
$a^{\dag}_{-\vec{q}}
(\vec{k}) = a^{\dag}_{\vec{q}}(\vec{k})$ and therefore
$u_{\vec{q}}(\vec{k})=u_{-\vec{q}}(\vec{k})$. Using this property and
the commutation relation (9) we find
\begin{equation}
U^{-1}(\vec{k}) \, a_{\vec{q}}(\vec{k}) \, U(\vec{k}) =
a_{\vec{q}}(\vec{k}) + \delta(\mu-\epsilon_{\vec{k}}) v_{\vec{k}}
u_{\vec{q}}(\vec{k})
\end{equation}
which, together with (8), leads us to the eigenvalue equation,
\begin{equation}
a_{\vec{q}}(\vec{k}) \mid u_{\vec{q}}(\vec{k}) \rangle =
\delta(\mu-\epsilon_{\vec{k}}) \, v_{\vec{k}} \, u_{\vec{q}}(\vec{k})
\mid u_{\vec{q}}(\vec{k}) \rangle.
\end{equation}
It is easy to see that $u_{\vec{q}}(\vec{k})$ is the displacement
of the Fermi Surface at the point $\vec{k}$ in the direction of $\vec{q}$.
Indeed, suppose we change the shape of the Fermi surface at
some point $\vec{k}$ by an amount $u(\vec{k})$. The occupation number
changes up to leading order by $\delta \langle n_{0}(\vec{k})\rangle =
-\frac{\partial \langle n_{0}(\vec{k})\rangle}
{\partial \vec{k}} u(\vec{k}) = v_{\vec{k}} \delta(\mu-\epsilon_{\vec{k}})
u(\vec{k})$ which is precisely the quantity which appears in (13).
Hence, the coherent states of eq~(10) represent {\it deformed Fermi
Surfaces} parametrized by the bosonic field $u_{\vec{q}}(\vec{k})$.

Next we define a many-body state which is a direct product of
the coherent states defined above,
\begin{equation}
\mid \{u\} \rangle = \prod_{\vec{k}} \otimes U(\vec{k}) \mid FS \rangle =
\Xi[u] \mid FS \rangle
\end{equation}
where, due to the commutation relation at different $\vec{k}$'s,
\begin{equation}
\Xi[u] = \exp\left(\sum_{\vec{k},\vec{q}} \frac{v_{\vec{k}}}
{2 \mid \vec{q}.\vec{v}_{\vec{k}} \mid} \, u_{\vec{q}}(\vec{k}) \,
a^{\dag}_{\vec{q}}(\vec{k})\right).
\end{equation}
The adjoint is simply,
\begin{equation}
\Xi^{\dag}[u] = \exp\left(\sum_{\vec{k},\vec{q}} \frac{v_{\vec{k}}}
{2 \mid \vec{q}.\vec{v}_{\vec{k}} \mid} \, u^{\ast}_{\vec{q}}(\vec{k}) \,
a_{\vec{q}}(\vec{k})\right).
\end{equation}

{}From the above equations we obtain the overlap of two of these
coherent states
\begin{equation}
\langle \{w\}\mid \{u\} \rangle = \langle FS \mid \Xi^{\dag}[w]
\Xi[u] \mid FS \rangle
= \exp\left(\sum_{\vec{k},\vec{q}} \frac{v_{\vec{k}}^2 \delta(\mu -
\epsilon_{\vec{k}}) }
{2 \mid \vec{q}.\vec{v}_{\vec{k}}\mid} \, w^{\ast}_{\vec{q}}(\vec{k})
 \, u_{\vec{q}}(\vec{k})\right) .
\end{equation}
It is also possible to find the resolution of the identity for this
Hilbert space,
\begin{eqnarray}
1 = \int...\int \prod_{\vec{k},\vec{q}} \left( \frac{v^2_{\vec{k}} \delta(\mu
-\epsilon_{\vec{k}})}{2 \pi i \, \mid \vec{q}.\vec{v}_{\vec{k}}\mid} \,
du_{\vec{q}}(\vec{k}) \, du^{\ast}_{\vec{q}}(\vec{k})
\mid u_{\vec{q}}(\vec{k}) \rangle \langle u_{\vec{q}}(\vec{k}) \mid \right)
\nonumber \\
\exp\left(-\sum_{\vec{k},\vec{q}} \frac{v_{\vec{k}}^2
\delta(\mu-\epsilon_ {\vec{k}})}
{2 \mid \vec{q}.\vec{v}_{\vec{k}}\mid} \, \mid u_{\vec{q}}(\vec{k})\mid^2
\right)
\end{eqnarray}
and we conclude, as expected, that they are overcomplete~\cite{klauder}.

In order to study the dynamics of these modes
we can construct, from (17) and (18), a generating functional as a sum
over the histories of the Fermi Surface in terms of
these coherent states in the form $Z = \int D^2u \, \exp\{i \, S[u]\}$ where
$S$ is the action whose Lagrangian density is given by ($\hbar=1$),
\begin{equation}
L[u] = \sum_{\vec{k},\vec{q}} \frac{v_{\vec{k}}^2
\delta(\mu-\epsilon_{\vec{k}})
}{\mid \vec{q}.\vec{v}_{\vec{k}}\mid} \, i \, u^{\ast}_{\vec{q}}(\vec{k},t)
\frac{\partial u_{\vec{q}}(\vec{k},t)}{\partial t} -
\frac{\langle \{u\}\mid H \mid \{u\} \rangle}{\langle \{u\}\mid \{u\} \rangle}
\end{equation}
where $H$ is the Hamiltonian of the system.

Consider now the the simplest Hamiltonian possible, for instance,
one which is quadratic in the operators $a^{\dag}$ and $a$,
\begin{equation}
H= \frac{1}{V} \sum_{\vec{k},\vec{k'},\vec{q}} G^{\vec{q}}_{\vec{k},\vec{k'}}
a^{\dag}_{\vec{q}}(\vec{k}) \, a_{\vec{q}}(\vec{k'}).
\end{equation}
where $V$ is the volume of the system.
Using (13) we write the Lagrangian density in the form
\begin{equation}
L[u] = \sum_{\vec{k},\vec{k'},\vec{q}} \frac{v_{\vec{k}}^2
\delta(\mu-\epsilon_{\vec{k}})}{\mid \vec{q}.\vec{v}_{\vec{k}}\mid}
\, u^{\ast}_{\vec{q}}(\vec{k},t) \left(i \delta_{\vec{k},\vec{k'}}
\frac{\partial}{\partial t} - \frac{1}{V} \, W^{\vec{q}}_{\vec{k},\vec{k'}}
\delta(\mu-\epsilon_{\vec{k'}})\right) u_{\vec{q}}(\vec{k'},t)
\end{equation}
where
\begin{equation}
W^{\vec{q}}_{\vec{k},\vec{k'}}= \frac{\mid \vec{q}.\vec{v}_{\vec{k}}\mid
\, v_{\vec{k'}}}{v_{\vec{k}}} G^{\vec{q}}_{\vec{k},\vec{k'}}.
\end{equation}
Notice that this action is quadratic in the fields and, therefore, the
semiclassical approximation is exact. The semiclassical equations of
motion for this action
are given by the saddle point equation derived from $L$, namely,
\begin{equation}
i \frac{\partial u_{\vec{q}}(\vec{k},t)}{\partial t} = \frac{1}{V}
\sum_{\vec{k'}} W^{\vec{q}}_{\vec{k},\vec{k'}} \delta(\mu-\epsilon_{\vec{k'}})
u_{\vec{q}}(\vec{k'},t).
\end{equation}
is a partial integral equation. Observe that the vectors $\vec{k}$ are
at the Fermi Surface and, therefore, the above equation only depends on
the solid angle, $\Omega$,  and the velocity $\vec{v}_{\vec{k}}$.
We can therefore integrate over the magnitude of $\vec{k'}$ and use that the
density of states at the Fermi Surface is given by,
$N(0) = \sum_{\vec{k}} \delta(\mu-\epsilon_{\vec{k}})$.
Therefore, in the thermodynamic limit we replace (23) by,
\begin{equation}
i \frac{\partial u_{\vec{q}}(\Omega,t)}{\partial t} = \frac{N(0)}{S_d}
\int d\Omega' \, W^{\vec{q}}_{\Omega,\Omega'} \, u_{\vec{q}}(\Omega',t).
\end{equation}
where $S_d = (2 \pi)^{-d} \int d\Omega$. In particular, we choose,
\begin{equation}
G^{\vec{q}}_{\vec{k},\vec{k'}} = \frac{\delta_{\vec{k},\vec{k'}}}{N(0)}
\, + f_{\vec{k},\vec{k'}}
\end{equation}
and define the angle $\theta$ such that
$\vec{q}.\vec{v}_{\vec{k}} = q v_F \, cos \theta$. With the choice of eq~(25)
 the ${\vec q}=0$ ambiguity mentioned above is removed.
It is easy to see that equation (24) becomes
\begin{equation}
i \frac{\partial u_{\vec{q}}(\Omega,t)}{\partial t} =
q v_F \, cos \theta \, u_{\vec{q}}(\Omega,t) + q v_F \, cos \theta
\int d\Omega' \, F_{\Omega,\Omega'} \, u_{\vec{q}}(\Omega',t)
\end{equation}
with
$F_{\vec{k},\vec{k'}} = N(0) f_{\vec{k},\vec{k'}}.$
Eq~(26) is the Landau equation of motion
for sound waves (the collective modes) of a neutral Fermi Liquid
where $f_{\vec{k},\vec{k'}}$ is the scattering amplitude for
particle-hole pairs ~\cite{baym}.

To understand the reason for this result we rewrite the Hamiltonian $H$ in
terms of the original operators (2). Using (6) and (7) we find,
\begin{equation}
H= \frac{1}{V} \sum_{\vec{k},\vec{k'},\vec{q}} G^{\vec{q}}_{\vec{k},\vec{k'}}
n^{\dag}_{-\vec{q}}(\vec{k}) \, n_{\vec{q}}(\vec{k'}) \Theta(sgn(q)).
\end{equation}
which gives the the change in the energy up to second order in the deviations
$\delta n(\vec{k})$ of the Landau theory of the Fermi Liquid when the Fermi
Surface changes from $\vec{k}_F$ to $\vec{k}_F + \delta \vec{k}_F(\theta)$.
Observe that this Hamiltonian is interacting in terms of the original
electrons. Recall that in one dimension the bosonized theory is also
quadratic (in the absence of relevant operators).

We conclude therefore that our bosons represent sound waves which
propagate around the Fermi Surface at zero temperature. The solution of
eq~(26) will give the
possible values for the frequencies of oscillation for these modes and
they will depend essentially on the Landau parameters of the theory.
The Landau equation, eq~(26), yields solutions which represents both
stable collective modes (``sounds") as well as solutions with imaginary
frequency which represent the decay of these modes in the particle-hole
continuum.
The presence of this second type of solutions implies that the path
integration must also be done over unstable configurations. This
behavior is a direct consequence of the phase space. Notice that in one
dimension these unstable solutions are absent and only the collective
mode is left.

In summary, in this paper we have shown that it is possible
to bosonize the particle-hole
elementary excitations of a Fermi Liquid in any number of dimensions
consistently with the Landau Theory.
We have shown that the bosons are coherent superposition of
particle-hole pairs close to the Fermi Surface. We also have shown
that the coherent states associated with these excitations are
related with sound waves which propagate in the Fermi Surface.
Our approach gives for simple Hamiltonian models, in the semiclassical
limit, the Landau equation which describes the propagation of these
sound waves.

One of us (A.H.C.N.) acknowledges J.~M.~P.~ Carmelo for useful
discussions and CNPq (Brazil) for a scholarship.
This work was supported in part by NSF Grant No. DMR91-22385 at the
University of Illinois at Urbana-Champaign.

\end{document}